\documentclass[pre,aps,twocolumn,
pdflatex,
superscriptaddress,floats,showpacs,10pt]{revtex4-1}
\usepackage{graphicx}
\usepackage{dcolumn}
\usepackage{bm}
\usepackage{amssymb}
\usepackage{amsmath}
\usepackage{textcomp}
\usepackage{color}

\begin{document}

\title{Laser Ion Acceleration from Mass-Limited Targets with Preplasma}

\author{K. V. Lezhnin}
\affiliation{Institute of Physics of the ASCR, ELI-Beamlines, Na Slovance 2, 18221 Prague, Czech Republic}
\affiliation{Moscow Institute of Physics and Technology, 
Institutskiy per. 9, Dolgoprudny, Moscow Region 141700, Russia}

\author{F. F. Kamenets}
\affiliation{Moscow Institute of Physics and Technology, 
Institutskiy per. 9, Dolgoprudny, Moscow Region 141700, Russia}

\author{T. Zh. Esirkepov}
\affiliation{Kansai Photon Science Institute, 8-1-7 Umemidai, Kizugawa, Kyoto 619-0215, Japan}

\author{S. V. Bulanov}
\affiliation{Kansai Photon Science Institute, 8-1-7 Umemidai, Kizugawa, Kyoto 619-0215, Japan}
\affiliation{A. M. Prokhorov General Phys. Inst. of RAS, Vavilov Str. 38, Moscow 119991, Russia}

\author{O. Klimo}
\affiliation{Institute of Physics of the ASCR, ELI-Beamlines, Na Slovance 2, 18221 Prague, Czech Republic}
\affiliation{Faculty of Nuclear Sciences and Physical Engineering, Czech Technical University in Prague, 11519 Prague, Czech Republic}
\author{S. Weber}
\affiliation{Institute of Physics of the ASCR, ELI-Beamlines, Na Slovance 2, 18221 Prague, Czech Republic}
\author{G. Korn}
\affiliation{Institute of Physics of the ASCR, ELI-Beamlines, Na Slovance 2, 18221 Prague, Czech Republic}

\date{\today}

\begin{abstract}
The  interaction of high intensity laser radiation with mass-limited target
exhibits significant enhancement of the ion acceleration
when the target is surrounded by an underdense plasma corona,
as seen in numerical simulations.
The self-generated quasistatic magnetic field 
squeezes the corona causing the intensification of
a subsequent Coulomb explosion of the target.
The electric field intensification at the target edges
and plasma resonance effects 
result in the generation of characteristic density holes
and further contributes to the ion acceleration. 

\bigskip

\end{abstract}

\maketitle

\section{Introduction}

High energy ion generation in the interaction of 
intense laser pulses with mass-limited targets
is promising for applications in a wide range,
from the development of ion sources for medicine and
the fast ignition of controlled thermonuclear fusion
to 
the investigation of warm dense matter, high energy density phenomena,
and laboratory astrophysics
(see review articles \cite{IonsREV, LED, HTUFN, MIRR, RelAstro} and references therein). 
Mass-limited targets otherwise called reduced mass targets
have a finite transverse size comparable with the laser focal spot diameter
\cite{KLUGE,ZIGLER,Zeil,ZIGLER2}.
Their principal advantage
is that an intense laser pulse can remove much more electrons from it
than from a wide and thick target.
This generates a greater electrostatic potential thus enhancing the ion acceleration. 
In wider targets, electrons from the periphery additionally reduce
the ion acceleration quickly  smoothing out the electric potential,
while in thick targets, the laser radiation cannot reach deeper layers.

The advantage of mass-limited targets is best seen with isolated clusters,
from which an intense laser sweeps all the electrons.
Then the Coulomb explosion occurs:
the ions are accelerated under the repulsion force of uncompensated electric charge.
A pure Coulomb explosion provides an isotropic ion acceleration
with very characteristic energy spectrum, where a large number of ions acquire a high energy
\cite{CEL}.

When the laser radiation can not penetrate inside the target deeper than the skin depth,
it heats electrons creating a plasma sheath with a strong electric charge separation.
The latter causes the ion acceleration
in the target normal sheath acceleration (TNSA) regime \cite{TNSAG, TNSAW, TNSAM}.
In mass-limited targets, this regime can give a substantial contribution
to the ion acceleration
when the laser pulse power is far from being enough to wipe all the electrons.

For sufficiently high intense laser pulses and optimally dense targets,
the laser radiation pressure pushes the target as a whole in the propagation direction,
while the target remains, on average, mostly quasi-neutral.
This is the regime of the radiation pressure dominated acceleration (RPDA) of ions
\cite{RPDAV, RPDAE, RPDAK, SSB-guid}.
A tailored laser pulse can provide in principle unlimited ion acceleration \cite{UNLIM}.
In the case of a mass-limited target, the RPDA of ions can be stable
even if the target is initially off-axis \cite{LEZHNIN}. 
A combination of the RPDA and a subsequent Coulomb explosion of a thick target
leads to a directed Coulomb explosion regime of the ion acceleration \cite{DCE}.


In general, a laser system produces a high-intensity short-duration (main) pulse on top of a relatively low-intensity nanosecond (background) amplified spontaneous emission (ASE), possibly with a few prepulses.
The main-pulse-to-background intensity ratio is called the laser pulse contrast.
When a laser pulse with a finite contrast irradiates a solid target,
preplasma is created before the main pulse arrival.
Preplasma is a plasma region where the density
gradually drops from values typical to solid state 
at some depth in the target down to
a value well below the critical density. 
Preplasma created around structured snow-targets irradiated by intense laser pulses
facilitates the efficient ion acceleration via the edge field intensification effect \cite{ZIGLER,ZIGLER2}.
The main pulse interacting first with preplasma exhibits regimes typical to 
gaseous targets, especially when the plasma density is near-critical.
Under these conditions, laser radiation can form a long-living quasi-static magnetic dipole
leading to the Magnetic Vortex Acceleration regime \cite{KUZNETSOV,Fukuda2009},
or can create a shock wave putting into effect the Shock Wave Acceleration \cite{SHOCK}.
While the main pulse loses its energy in preplasma it can also 
self-focus due to relativistic effects \cite{Rel-SF},
thus an optimised preplasma can crucially enhance the ion acceleration \cite{PREPLASMA}. 


Here we investigate how the above mentioned ion acceleration mechanisms 
reveals itself depending on the geometry of a mass-limited target 
and a plasma corona around it.
We carry out two-dimensional (2D) Particle-in-Cell (PIC) simulations
using the REMP code
based on the density decomposition scheme \cite{REMP}.
We describe the effects occurring in the presence of a plasma corona
around a mass-limited target: 
the enhancement of the Coulomb explosion of the ion core of the target
due to the density squeezing by the quasistatic magnetic field,
formation of a density hole near the tip of the target due to plasma resonance,
etc.
The paper is organized as follows.
In the next Section, we describe the laser and target parameters and the simulation configuration.
In Section III, the results of PIC simulations of the 
laser pulse interaction with mass-limited targets 
in a wide range of the irradiation parameters are presented.
There we discuss the effects of magnetic squeezing and 
electric field intensification at the target edge on the ion acceleration.
In Section IV, we consider the high power limit 
when the ion acceleration occurs in the radiation pressure dominated acceleration regime.
In the concluding Section, we summarize the results obtained.

\section{Simulation configuration}

\begin{figure}
\includegraphics[width=8cm]{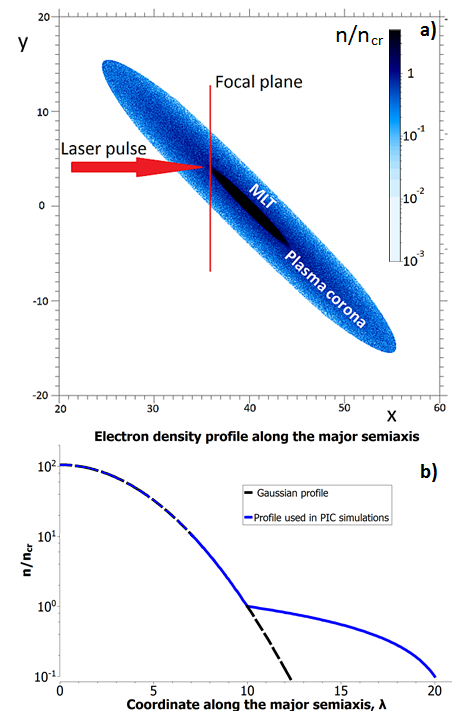}
\caption{a) A configuration of the laser interaction with the mass-limited target {(MLT)},
the electron density, the  laser propagation direction and the focal plane. Hereafter, all spatial dimentons are normalized to laser wavelength $\lambda$, temporal one - to laser period $2 \pi / \omega_0$, densities are normalized to critical density $\rm n_{cr}$, and electromagnetic fields are normalized to $\rm E_0 = m_e \omega_0 c / e$. 
b) The electron density profile along the major semiaxis.
Hereafter we call the Gaussian profile as ``shortened'' 
and complex one marked with blue as ``extended''.
}
\label{fig1}
\end{figure}

In order to investigate the processes that occur during 
the interaction of a relativistically strong laser pulse with 
a mass-limited water ice target
surrounded by a plasma corona, 
we conduct a series of 2D PIC simulations. 
The laser pulse is obliquely incident on the target
as is shown in the upper panel of Fig. \ref{fig1}. 

The plasma corona with a typical size of the order of several tens of micrometers plays an important role in the laser interaction with mass-limited targets \cite{KLUGE}. 
When it is created due to a finite contrast of the laser pulse,
it has a typical density profile consisting of three regions (see also Ref. \cite{PREPLASMA}). 
The first region is a plateau with the solid density,
corresponding to unperturbed portion of the target.
The second region has the density profile
which can be approximated by the Gaussian function
as shown in the lower panel of Fig. \ref{fig1}.
There the density decreases from the solid density value to the
critical density, $n_{cr} = \pi/r_e \lambda^2$,
corresponding to the laser wavelength $\lambda$.
Here $r_e = e^2/m_e c^2 \approx 2.8\times 10^{-13}\,$cm is classical electron radius.
In the third region, the plasma density linearly drops with a substantially larger scale.

In our simulations, the target has the shape of an ellipsoid,
so that the same density value curves are {ellipses}.
For such a shape
it is enough to describe the density profile along the major semi-axes.
We assume the following density profile.
The density decreases from maximum
according to the Gaussian function
down to $n_{cr}$, then it linearly drops down to $0.1 n_{cr}$,
then it is truncated to zero.
In order to restrict the total number of quasi-particles
representing plasma
(which amounts to $1.2\times 10^6$),
we limit the plasma density by the cutoff at $10 n_{\rm cr}$.
For moderate laser intensities,
when the laser radiation pressure is not dominant in the interaction,
this artificial cutoff does not significantly affect the simulation results,
as justified by a number of test simulations performed without a density cutoff
(note the analogous computational trick used in Ref. \cite{KLUGE}).

Here we present the results for a relatively high eccentricity, $\epsilon=0.995$,
of the ellipsoid representing the target shape.
This choice allows emphasizing the effects of 
a quasistatic magnetic field generated during the interaction.
In preliminary simulations (not shown here) 
we have found that 
for targets with a smaller eccentricity 
the quasi-static magnetic field is not sufficiently symmetrical
with respect to the target major axis.
The ion energy reaches maximum when the magnetic field is generated in a symmetrical way
(this is applicable for moderate laser intensities, 
when the laser radiation pressure does not dominate the interaction).
If the target has the form of a foil 
(in terms of ellipsoid target, it means that the major semiaxis 
is substantially larger than the laser focal spot size), 
the electrons move away from the region of irradiation,
along the target surface,
so that the associated quasi-static magnetic field
disperses along the target surface and eventually dissipates.


The laser pulse has a super-Gaussian profile with an index of 4,
so that the laser pulse intensity time profile
is described by the function
$I = 1.37\times 10^{18} \times a_0^2 (2)^{-(2t/\tau)^4}
\times (1{\rm\mu m}/\lambda)^2 \,$W/cm$^2$. The transverse laser pulse profile is also super-Gaussian with an index of 4.
The dimensionless amplitude, $a_0=eE_0/m_e \omega c$,
characterizes the laser electric field stregth, $E_0$.
Here $\omega=2\pi c/\lambda$.
The laser pulse duration is $\tau = 15 \lambda/c$,
its focal spot diameter is $5 \lambda$.
The laser pulse is p-polarized, i.e., 
its electric field vector is in the simulation plane $(x,y)$.
In our simulations, the laser amplitude
varies from $a_0=3$ to $a_0=134$,
which corresponds to the peak laser power in the range from 5 TW to 10 PW.
The laser focal plane is at the sharp edge of the target
where the density is $n_e=n_{\rm cr}$, if not stated otherwise.
The laser pulse is obliquely incident on the target at the angle of 45 degrees.
This configuration is chosen to maximize the laser energy transmission to the target.
For references concerning the dimensional values of the target density,
laser energy and intensity, etc,
the laser wavelength can be assumed to be $\lambda = 0.8 \mu m$.
For the scaling into the high laser power, 
we used the same profile but multiplied by 
some factor in accordance with the 
$a_0= \pi n_e l_0 /n_{cr} \lambda$ criteria, 
where $l_0$ is the target thickness
(the size along the minor axis). 

Our target is composed of electrons, protons, and oxygen ions
corresponding to frozen water,  $\rm H_2 O$. The follwoing mass ratios has been used: $\rm m_p / m_e =1836$, $\rm m_O / m_p =16$.
As is well known, one of the parameters that could influence 
the ion acceleration significantly 
is the ionization degree, $\rm Z$, 
and its spatial distribution \cite{IONIZE}.
For the sake of simplicity, 
in our simulations we assume a homogeneous spatial distribution of the ionization degree, which is fixed during the simulation's run and does not exceed the maximum value of $\rm Z_{\rm max}$. 
We estimate this ionization degree maximum
within the framework of the optical field ionization model.
Following Ref. \cite{IONIZE},
we obtain ${\rm Z}_{\rm max}^{2.4} \approx C_a \times a_0$,
with $C_a\sim 12$.
Depending on the laser amplitude,
in our simulations we initially set oxygen ions
with the ionization degree of +3, +5, +7 and +8.

In our simulations,
the grid mesh size is $\lambda/32$ in each spatial direction.
For moderate laser intesities, 
the simulation box has the size of $80\lambda \times 120 \lambda$. 
For higher laser intensity, $a_0 > 20$,
the simulation box is larger, $200\lambda \times 200 \lambda$),
in order to avoid stronger unphysical effects from the boundaries.
The boundaries are absorbing for particles. 
For the electromagnetic radiation, the boundaries
are periodical.
%
The space and time values are given in units 
of the laser wavelength and period,
i.e., $\lambda$ and $2 \pi /\omega$, respectively.

\section{Simulation Results}

We start our consideration with the case of a $5$ TW laser pulse,
corresponding to $a_0=3$.
While such the laser power can be considered 
relatively low in comparison with the up-to-date laser systems, 
it allows us to clarify a number of effects that can occur in the high-power limit.

\begin{figure}
\includegraphics[width=8 cm]{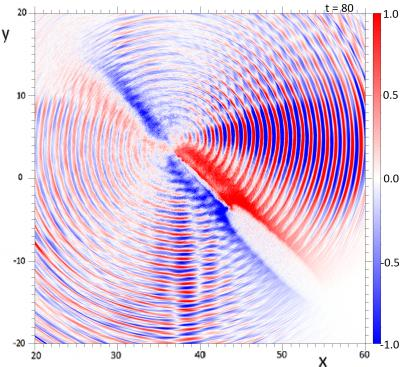}
\caption{Magnetic field distribution in the $(x,y)$ plane at $t=80$ for $a_0=3$ case
showing the laser pulse scattering at the target tip and 
reflection by the overdense plasma region.}
\label{cylwave}
\end{figure}

Fig. \ref{cylwave} shows the magnetic field distribution
at a final stage of the laser pulse interaction with the target.
About $60 \%$ of the laser pulse energy passes through,
because of the location of the focus at the target tip,
where plasma is underdense.
Nevertheless, a substantial part of the laser energy is absorbed:
eventually about $35 \%$ is transferred to the particle energy.
One of the absorption mechanisms is 
the plasma resonant absorption,
because for a p-polarized electromagnetic wave
obliquely incident on the target {the projection of 
the vector of electric field on the plasma density gradient 
does not vanishes}, $\vec{E} \cdot \nabla n \neq 0$ 
(for detals of the  plasma resonance phenomenon see, e.g., Refs. \cite{VLG, BKS}). 
A small part of the laser pulse is scattered at the target tip and 
specularly reflected by the overdense plasma region, as is clearly seen in Fig. \ref{cylwave}.
Scattered radiation has a form a cylindrical electromagnetic wave 
with an inhomogeneous distribution of the field amplitude. 
The transverse modulations seen in the transmitted and reflected pulses
are due to strong plasma density modulations
created by the incident laser pulse.

During the laser pulse interaction with the plasma corona, 
it undergoes relativistic self-focusing,
which can be seen in a slight increase of the transmitted pulse amplitude. 
We also note that 
the electric field strength at the target tip
increases due to the edge intensification of electromagnetic field
and the rise of plasma resonance \cite{Askar}.

A well-pronounced temporal periodicity of the maximum amplitude 
field (with the frequency close to that of the laser)
and 
the absence of substantial field enhancement in the
case of s-polarized pulse 
indicate the effect of the plasma resonance at the tip of the overdense target. 

Considering the simulation results for different initial laser amplitude $a_0$,
we find the following properties of the instantaneous 
electric and magnetic field strength maximum, $E_i$ and $B_i$, respectively.
For $a_0 < 10$,
$B_i\approx 2 a_0$, which is merely due to 
a transient standing wave formed during the laser pulse reflection 
from the target {and due to just logarithmic enhancement of the 
magnetic field in the resonance region (see Ref. \cite{LL8})}. Weaker reflection is observed for $a_0\ge 10$,
correspondingly, in this range $B_i\le 1.5 a_0$.
The dependence of the instantaneous electric field stregth maximum
on the laser amplitude changes from
$E_i\approx 6 a_0$ at $a_0 = 1$
to
$E_i\approx 2 a_0$ at $a_0 \ge 10$.

\subsection{Magnetic Squeezing}

\begin{figure}
\includegraphics[width=8cm]{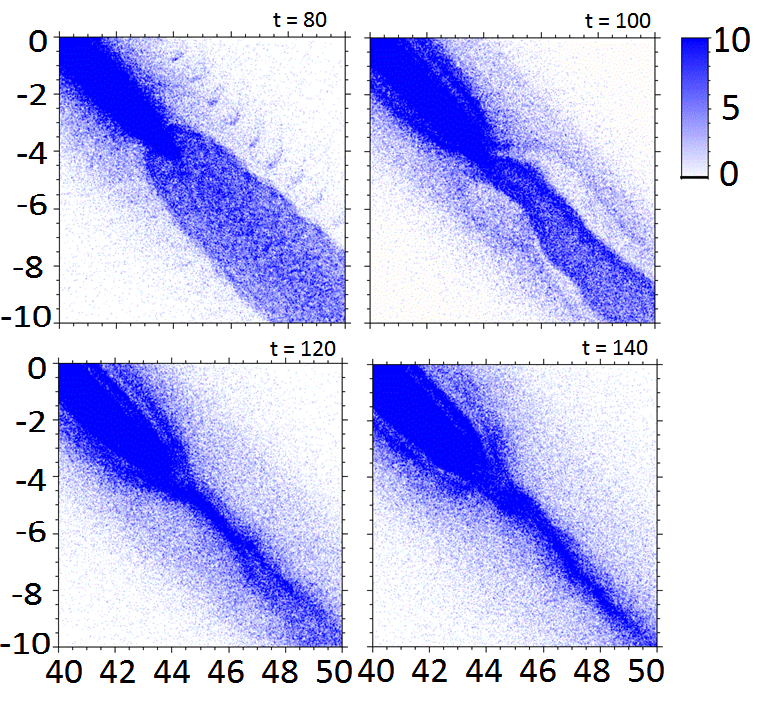}
\caption{
$a_0 = 3$. Electron density distribution at $t=80$, 100, 120 and 140. 
The underdense corona squeezing can be seen. 
Apart from that, the overdense part of the target is 
being shaped by the magnetic field.}
\label{bottomimpl}
\end{figure}

Fig. \ref{bottomimpl} shows the evolution of the electron density distribution
at the bottom part of the target, quite far from the laser pulse focus location,
for the case of $a_0=3$.
At $t=80\times 2\pi/\omega$,
the density modulations 
with the wavelength equal to that of the laser
are seen along the target.
At later time, the plasma corona is squeezed.
Eventually a high-density filament is formed,
as seen at $t=140 \times 2\pi/\omega$.
The sqeezing takes place for every kind of particles on their own timescale
determined by their charge-to-mass ratio. 

\begin{figure}
\includegraphics[width=8 cm]{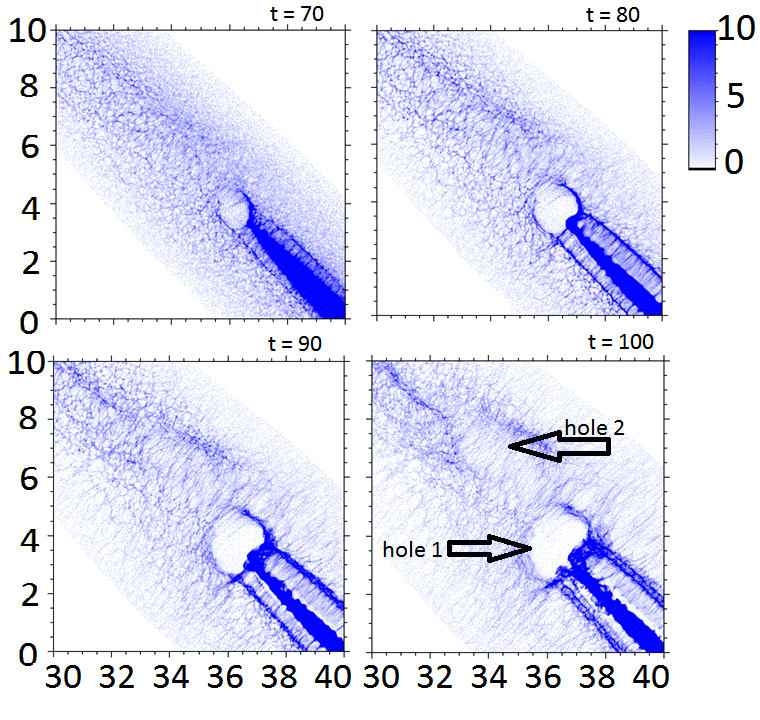}
\caption{
$a_0 = 3$. Proton density distribution at $t=70$, 80, 90 and 100. 
The hole appearance in the upper corona regions can be seen. 
The overdense part of the target is being compressed by the magnetic field.}
\label{upimpl}
\end{figure}

We also observe the appearance of the proton depletion regions 
in the upper part of the corona (see Fig. \ref{upimpl}), 
where the quasistatic magnetic field is maintained for a long time, 
making a substantial impact on the ion acceleration as well 
(hole 2, see Fig. \ref{upimpl}  and  \ref{magnup}).

\begin{figure}
\includegraphics[width=8cm]{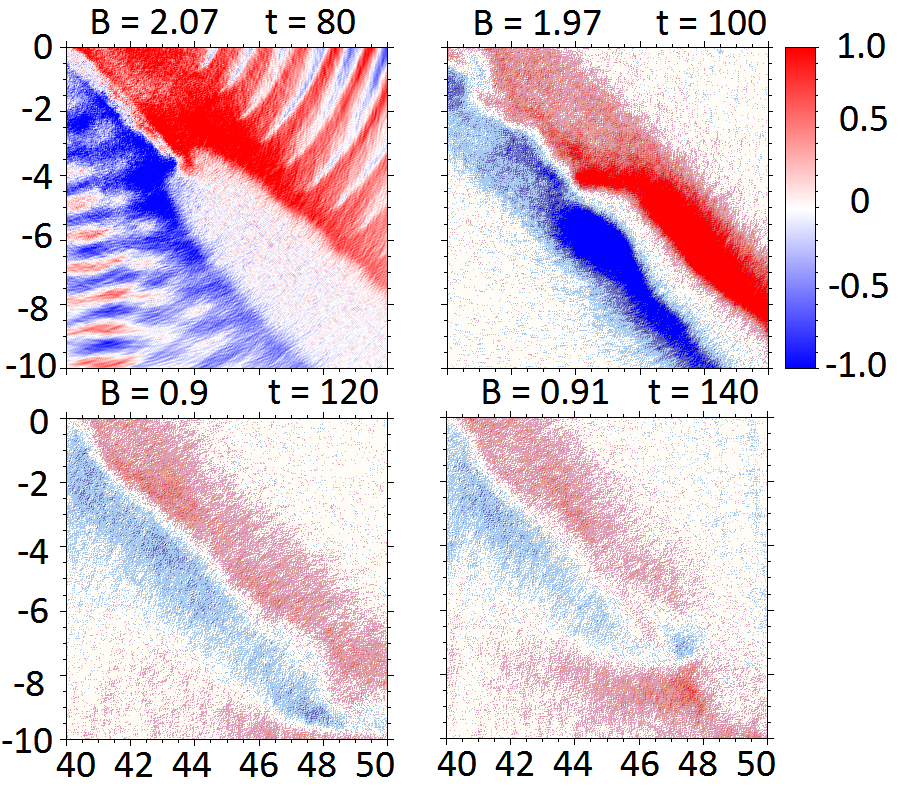}
\caption{
$a_0 =3$. Distribution of the $z$-component of the 
magnetic field in the lower target region at $t=80$, 100, 120 and 140. 
The appearance of the quasistatic magnetic field corresponds to the hole in proton density, 
which can be explained by the vortices propagating along the major semi-axis, 
compressing the target and then slowly decreasing due to the electron expansion to vacuum.}
\label{magnbottom}
\end{figure}

\begin{figure}
\includegraphics[width=8cm]{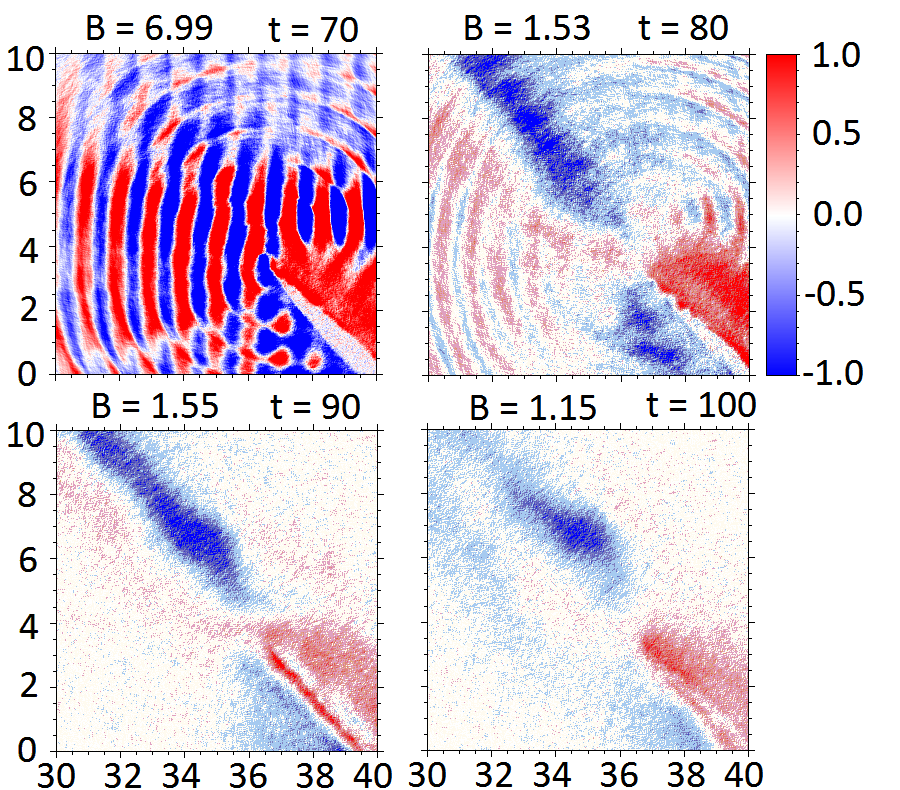}
\caption{
$a_0 =3$. Distribution of the $z$-component of the 
magnetic field in the upper target region at $t=70$, 80, 90 and 100. 
Vortices propagate along the major semi-axis, 
compress the target and then slowly decrease due to the electron expansion to vacuum.}
\label{magnup}
\end{figure}

Effect of the corona squeezing is strongly correlated 
with the magnetic field evolution. 
Figures \ref{magnbottom} and \ref{magnup} show the magnetic 
field evolution at the same instants of time. 
Laser pulse interacts with the electrons from the corona, 
forcing them to circulate around the upper and lower parts 
of the target in the $(x,y)$ plane. 
The electric current carried by the electrons generates the magnetic field, 
i.e., the magnetic vortex structure is being created. 
It is noticeable that the maximum value of the generated 
quasistatic magnetic field exceeds the maximum 
laser pulse amplitude by a factor of more than two. 
As is well known, the magnetic vortex formation is a 
common process in the laser irradiated underdense plasma \cite{VORTEX} having 
a density gradient \cite{KUZNETSOV}.
The crucial parameter for the vortices is the density gradient 
as it determines their lifetime and propagation velocity. 
 We have considered various configurations of the corona 
density distribution, and it turns out that density profile 
 should be gentle enough ($\rm \approx 0.1 n_{cr}/ \mu m$, 
 in agreement with \cite{KUZNETSOV}) 
in order to keep the vortices in the corona for relatively long time. 
However the plasma density gradient should not be too small because the 
vortices do not move in this case. 
 Moreover, the geometry of the underdense region affects the vortex formation dramatically. 
This is due to the strong connection 
 with the number of electrons trapped by the magnetic field of the pulse--the more electrons laser pulse sees during the propagation in the corona, 
 the more stable magnetic field is generated. 
Propagating along the major semiaxis of the ellipsoid, 
 these vortices force corona compression due to the magnetic 
pressure and then their amplitude gently decreases 
 due to the expansion of the electrons to vacuum (Fig. \ref{magnup}).

\begin{figure}
\includegraphics[width=8cm]{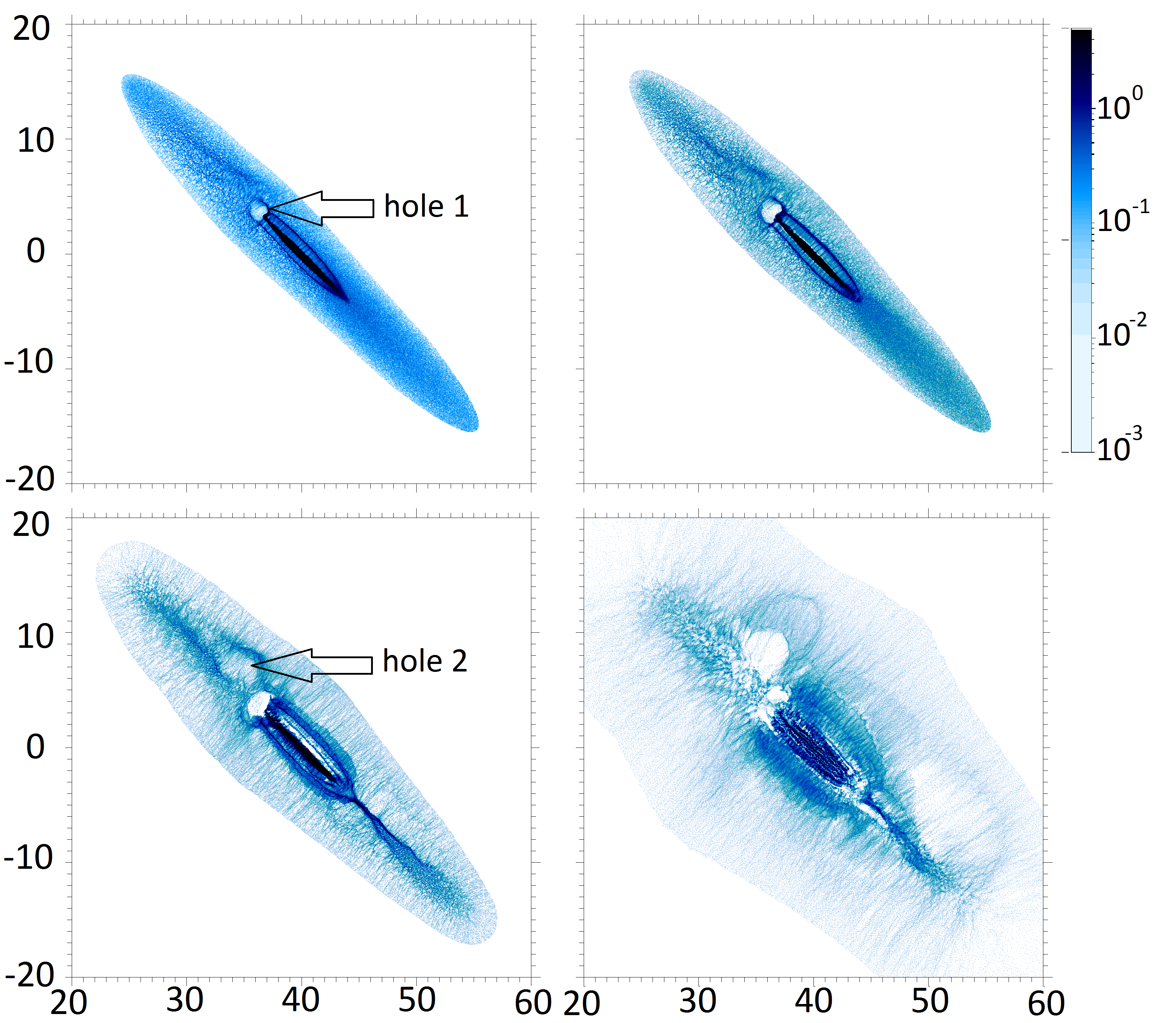}
\caption{
$a_0 = 3$. Distribution of the proton density at $t=80$, 100, 120 and 200. 
The hole boring in the upper corona regions and 
the corona squeezing in the bottom corona part can be seen.}
\label{fig5}
\end{figure}


\begin{figure}
\includegraphics[width=8cm]{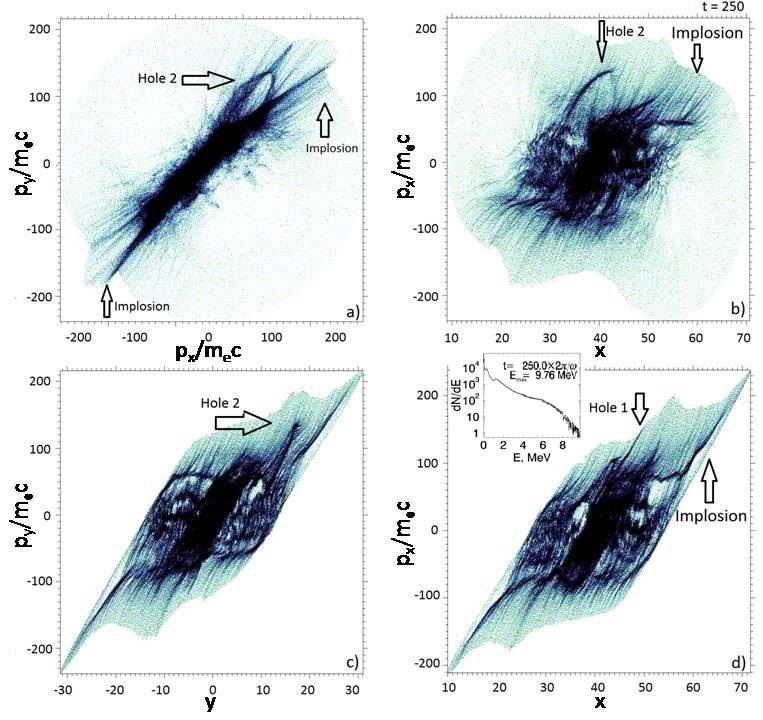}
\caption{
$a_0 = 3$. Proton phase space and the  energy spectrum ($(p_x,p_y)$, 
$(x,p_y)$, $(y,p_y)$, and $(x,p_x)$) for $t=250$ (a), b), c), and d), respectively);  protons accelerated from 
both holes can be seen along with the magnetic squeezing effect. Inset: proton energy spectrum, $\rm E_{max}\approx 10~MeV$.}
\label{fig8}
\end{figure}

\begin{figure}
\includegraphics[width=9 cm]{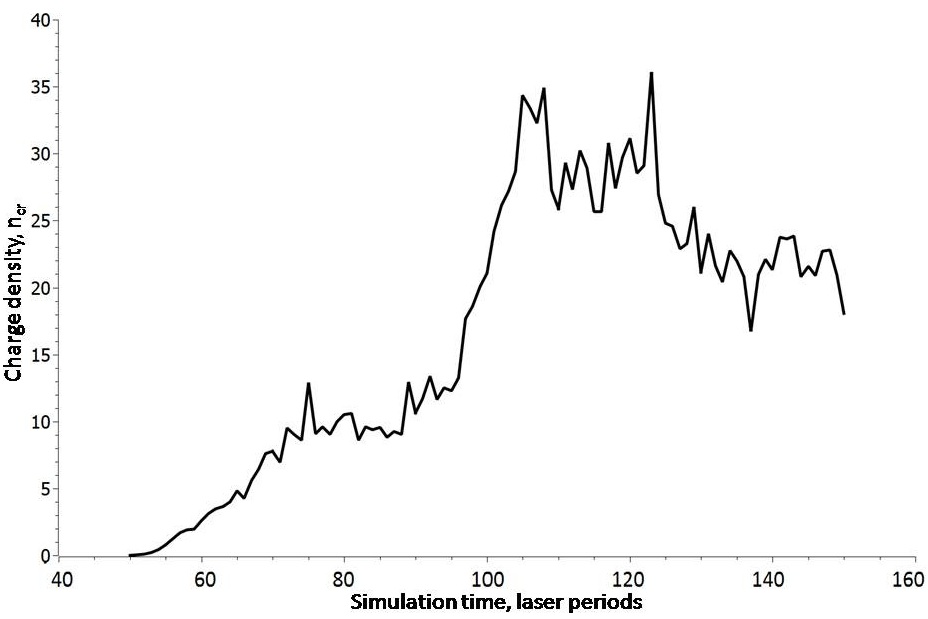}
\caption{
$a_0 = 3$. Maximum charge density distribution in 
bottom corona region ($44<x<50$, $-10<y<-4$, see Fig. \ref{bottomimpl}). 
We see a substantial 
increase of charge density at $t=120$, when the Coulomb explosion of 
this region initiates (see Fig. {fig5}) }
\label{chargedense}
\end{figure}

Proton component of the target experiences the squeezing on its own timescale. 
From initial ellipsoidal layer with the density below the critical one, 
it is being increased at least by a factor of 5, 
which leads to the more energetic Coulomb explosion than in more 
dense but also more quasi-neutral 
central part of the target.

We observe such effect 
in our simulations--we detect that for the case of $a_0 = 3$ 
we get at least by an order of magnitude larger positive-signed 
total charge density than we could possibly had by removing all the electrons 
from the the target without any pre-compression (see Fig. \ref{chargedense}). 
All the "hair'' with the largest energies in the configuration on the $(p_x,p_y)$ 
distribution in Fig. \ref{fig8} 
correspond to this type of acceleration.

Similar effect also takes place in the upper corona region (hole 2). 
Pre-accelerated by the magnetic field, it eventually experiences the Coulomb explosion. 
Finally, it gives one of the most energetic populations of 
accelerated protons in the system, see $(x,p_y)$ and $(y,p_y)$ phase planes in Fig. \ref{fig8}.

It is also notable that the magnetic field penetrates into 
the overdense target region (see Figs. \ref{magnbottom} $ and $ \ref{magnup}). 
As the solid state target part is always rich with electrons, 
the magnetic field would be sustained here for much longer time 
in comparison with corona vortices. These fields affect the ion acceleration process, 
boring the gaps and pushing the external target layers along the minor semi-axis, 
directing the acceleration of the protons (see Fig. \ref{fig5}).

Although there is an isotropic component of the proton momentum distribution 
(see Fig. \ref{fig8}, $(p_x,p_y)$ plot) that can originate from the Coulomb 
explosion of an external layer of the underdense corona, the most energetic particles 
(and also the largest fraction of protons) are distributed in the vicinity 
of the diagonal $p_y=p_x$ on a corresponding phase distribution plot (see Fig. \ref{fig8}). 
This strong anisotropy is in a tight connection with the geometry 
of the target having the form of a high eccentricity 
ellipsoid. The target can be regarded locally as a charged plane with 
the electric field orthogonal to the major semi-axis. 
However, there are some deviations from the main direction--filaments 
in the distribution of protons in the phase plane.
Some of the proton beams contain substantial number of particles compared 
with the isotropic component. For example, two filaments seen in the 
$(x,p_y)$ plot (the curved one around $x\approx 35-40$ and the plane one around $\approx 52-60$); 
even though they do not correspond to the maximum energy value, 
they are at least an order of magnitude more energetic than the particles in the 
central part of the target. 
Taking into account that the initial central density is 10 times 
larger than the maximum corona density, we can conclude that the magnetic 
squeezing does have a positive effect, providing larger initial 
density for the Coulomb explosion of the corona.

\subsection{Edge field intensification and plasma resonance}

\begin{figure}
\includegraphics[width=7cm]{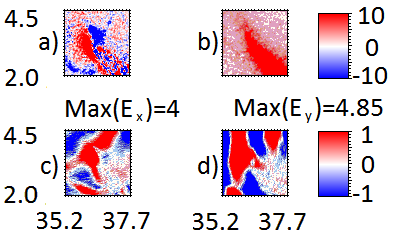}
\caption{a) charge density distribution around the target's tip; 
b) proton density distribution around the target's tip; 
c) and d): $E_x$ and $E_y$ electric field components. 
All graphs are presented at t=60, simulation for $a_0 = 3$ laser pulse. 
Strong charge separation, corresponding field enhancement and 
starting point of proton acceleration are seen.}
\label{resonance}
\end{figure}

\begin{figure}
\includegraphics[width=9cm]{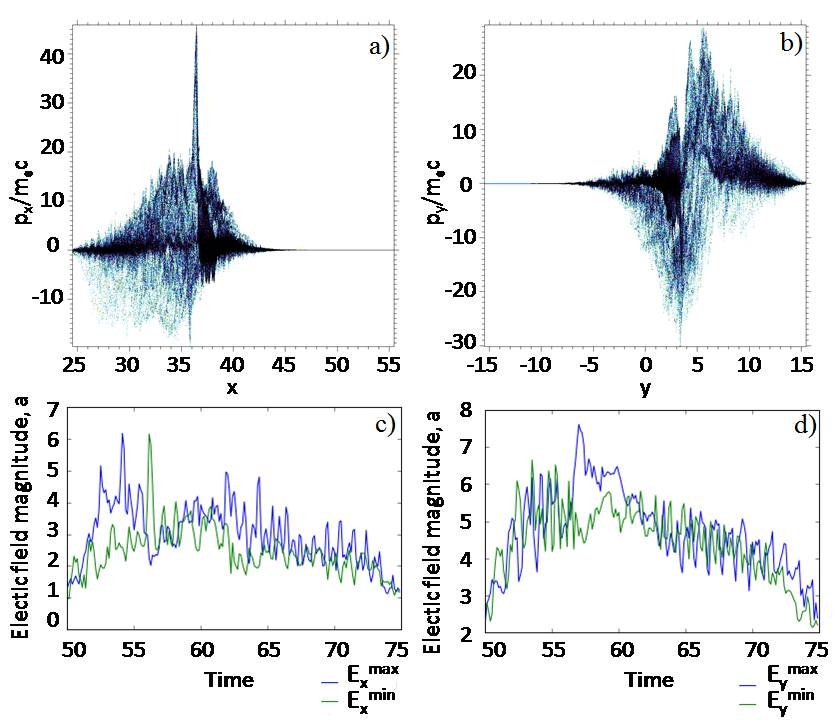}
\caption{
$a_0 = 3$. a) and b): proton phase distributions  in the planes 
$(x,p_x)$ and $(y,p_y$) for $t=60$; the peak of proton momentum magnitude at $(x=37, y=4)$ corresponds to plasma resonance at the sharp edge of the target. 
c) and d): distribution of electric field $E_x$ and $E_y$ projections amplitude; 
MAX and MIN denote to maximum positive and negative projections of 
the electric field in the resonant region (hole 1, see Figs. \ref{upimpl}  and \ref{fig5})}
\label{fig7}
\end{figure}

We note that we do not see substantially large magnetic fields in the vicinity of the target tip (hole 1). 
Fig. \ref{fig5} shows how the proton density evolves for 80, 100, 120 and 200 laser periods. 
As the laser pulse is focused on the sharp edge of the target, 
the electric field is amplified significantly 
in this region, which in its turn bores a hole in the proton density. 
The strong charge separation is observed at the sharp edge of 
the target for the whole period of laser propagation 
through the target, see Fig. \ref{resonance}. This hole might be 
associated with the plasma resonance \cite{RESONANCE} 
and edge field amplification \cite{Askar, LL8, KKM} which can lead 
to the particle acceleration from the target's edge and hole 1 formation 
(see Figs. \ref{resonance}  and  \ref{fig7}). 


Plasma resonance is attributed to the enhancement of the electric field 
in the region of critical plasma density. 
The maximum electric field amplitude in nonlinear Langmuir oscillations 
excited at the resonance surface is determined by the 
wave breaking process (see Ref. \cite{BKS} and literature cited therein). 
This results in the fast electron component generation and in the light ion 
acceleration. 
The light ion emission from the resonant region has been observed in our simulation. 
The light ion acceleration occurs  under action of 
the pondermotive force that arises in the high-frequency field localized in 
the plasma resonance region \cite{BKS, RESONANCE}. 
These ions are accelerated to the energies of the order of 
${\cal E}_i \approx Z {\cal E}_{E}$, where 
${\cal E_{E}}\approx m_e v_E^2/2=e^2 E_{m}^2/2 m_e \omega^2$, 
$E_m$ is the electric field in the plasma resonance region, 
corresponds to the electron quiver energy, and
$v_E$ is the electron quiver velocity. 
The proton energy, in our case, is approximately equal to
$a_{0}^2 m_e c^2 /4$ which is of the order of 
$0.5~\rm MeV$ for protons in case of $a_0=1$, corresponding to the electron quiver velocity. 
The ion phase distributions and charge density, proton density and projections 
of the electric field around the target tip can be seen in Fig. \ref{resonance}.

\begin{figure}
\includegraphics[width=6cm]{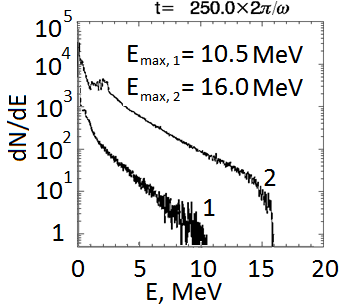}
\caption{Energy spectrum for flat-profile target (1) 
and inhomogeneous-profile target ("extended" corona, see Figure \ref{fig1}) setup (2), $a_0=3$.}
\label{energycompare}
\end{figure}

All these processes lead to the energy spectrum that is shown in Figure \ref{energycompare}. 
We present two spectra--one for the target discussed above 
and another with the flat density profile in order to demonstrate how 
the density gradient impacts the maximum ion energy. 
Targets have the same shape of ellipsoid and the flat target is 
limited by the ellipsoid with semiaxises of $\rm 1 \mu m \times 10 \mu m$. 
As we see, implementing targets with density gradients leads 
to the proton energy increase  up to $5 ~\rm MeV$. 
The maximum value of proton energy for corona-equipped 
target equals to $\approx \rm 16 MeV$, 
which can be considered as advanced acceleration level 
with the considered type of terawatt laser. 
The use of the corona-equipped target also provides better 
energy transmission, which is about $32 \%$ against $19 \%$ for the case of the flat target.

\subsection{High power limit}

Here we discuss the physical processes that dominate the 
laser-target interaction in the limit of higher peak laser power.
We show the results of the PIC simulations 
with the same target but with the larger laser amplitude $a_0$.
We take it equal to 20, which corresponds to $200~{\rm TW}$ laser power. 
In this high field amplitude limit, the light ion acceleration 
is less pronounced because of the nonlinear saturation of the resonance 
due to aperiodicity of plasma oscillations caused by the relativistic effects.

\begin{figure}
\includegraphics[width=8 cm]{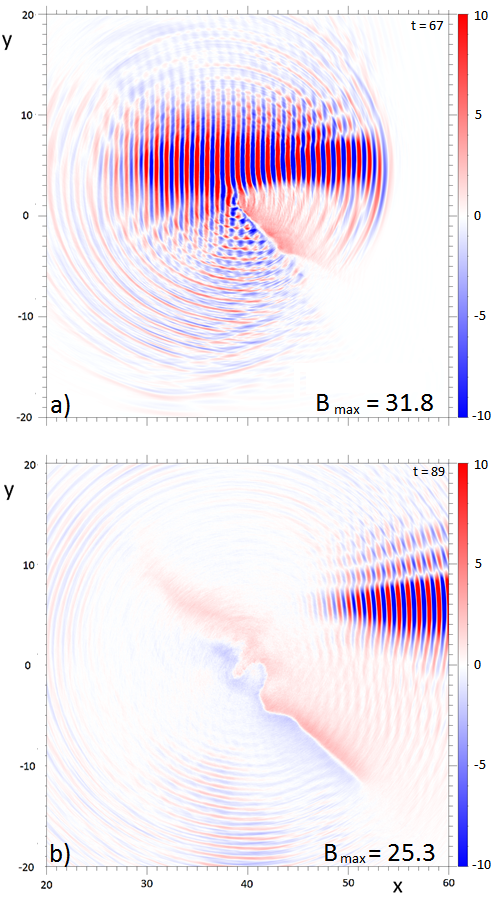}
\caption{Self-focusing for $a_0 = 20$ laser pulse in the plasma corona. a) The $z$-component of the magnetic field in the $(x,y)$-plane at $t=67$. 
b) The same as in the frame a) at $t=89$.}
\label{sf}
\end{figure}

Besides, the fraction of the radiation that passes through 
the target has the amplitude increased to about $\approx 20 \%$, 
which might be attributed to the self-focusing in the plasma corona, 
see Fig. \ref{sf}.

\begin{figure}
\includegraphics[width=8cm]{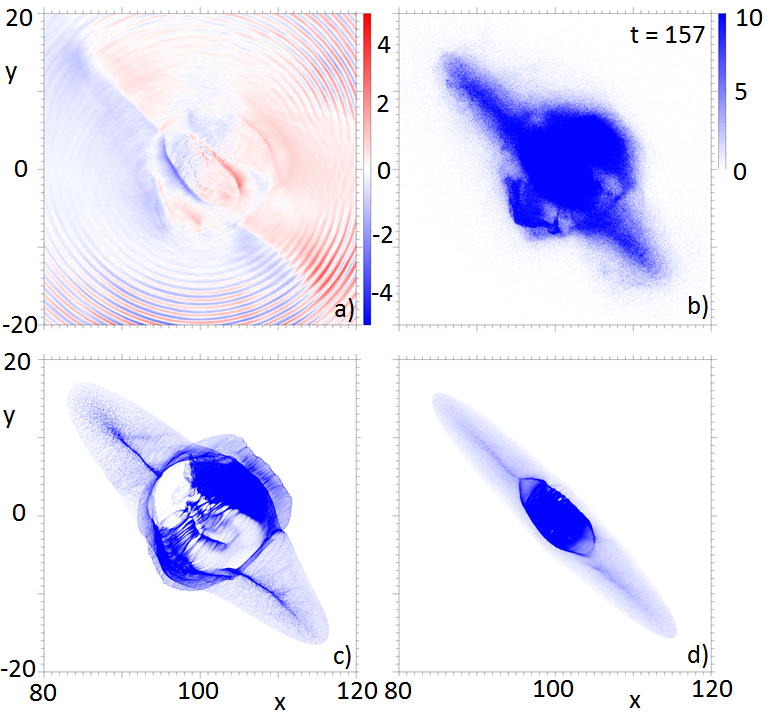}
\caption{a) The $x$-component of the electric field, 
b) The electron, proton (c) and oxygen (d) ion density distributions for $t=135$, $a_0 = 20$.}
\label{a20dens}
\end{figure}

\begin{figure}
\includegraphics[width=8cm]{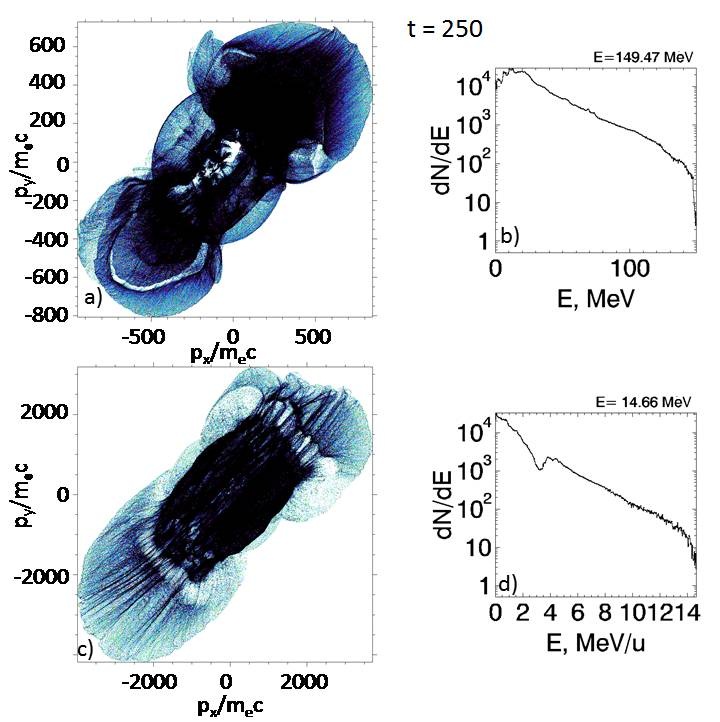}
\caption{The $(p_x,p_y)$ phase distributions of protons (a) and oxygen ions (c), energy distributions of protons (b) and oxygen ions (d), $a_0 =20$.}
\label{a20distr}
\end{figure}

The configuration of the target provides a high-level 
pulse energy transfer to the plasma and fast particle energy, energy transfer is about $52 \%$. 
We note that the energy transfer in the flat-density 
profile targets is lower, about $35 \%$.
During the interaction with the laser pulse, the electrons 
are being heated and expand into the vacuum, 
generating a strong charge separation along with the heavy oxygen 
ions with uncompensated charge that contribute 
to the repelling field, in which the protons are being accelerated. 
Ions are seen to move in two shell way (see Fig. \ref{a20dens}, $c$ and $d$).
The first shell of protons moves with higher speed and only afterwards, 
the shell of oxygen ions moves with lover velocity as seen in Fig. \ref{a20dens}. 
Corresponding electric field could be seen at $\rm E_x$ 
(the symmetrical picture can be seen for $\rm E_y$ as well) 
distribution figure as well - inner repelling fields are 
for oxygen shell and outer - for protons. 
In case of extended corona we observe the maximum energies of 
$160~ \rm MeV$ protons and $14~ \rm MeV/u$ 
oxygen ions with the angle and energy distributions presented in Figure \ref{a20distr}. 
With the exponential corona, we derived maximum energies at least $10 \%$ larger. 
Similar results have been obtained in Ref. \cite{ZIGLER2}.

As it has been shown in Ref. \cite{LEZHNIN}, 
the finite pointing precision of the laser imposes constraints 
on the maximum attainable energies 
of accelerated ions. 
However, while it is still true for the considered type of the target, 
the shift of the focus can give rise to a number of the effects 
that also provide energetic particles, even though their maximum energy is a factor of 1.5 smaller. 
For moderate laser power, we still observe that the most energetic particles form 
two opposite directing streams that are emitted from the target's tip. 
This is a consequence of both plasma resonance and edge field intensification, 
which give larger maximum field amplitudes.

\begin{figure}
\includegraphics[width=8cm]{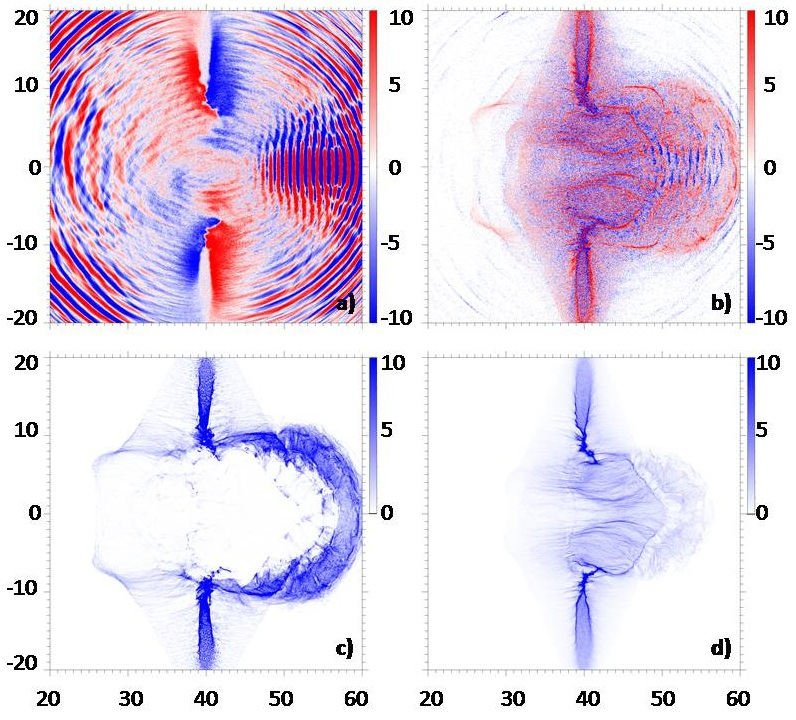}
\caption{The $z$-component of the magnetic field (a), 
the electric charge density (b), the proton (c) and the oxygen (d) 
density distributions in the $(x,y)$ plane for the case of 
the laser pulse normal incidence on the target;  $a_0=134$.}
\label{fig14}
\end{figure}

\begin{figure}
\includegraphics[width=8cm]{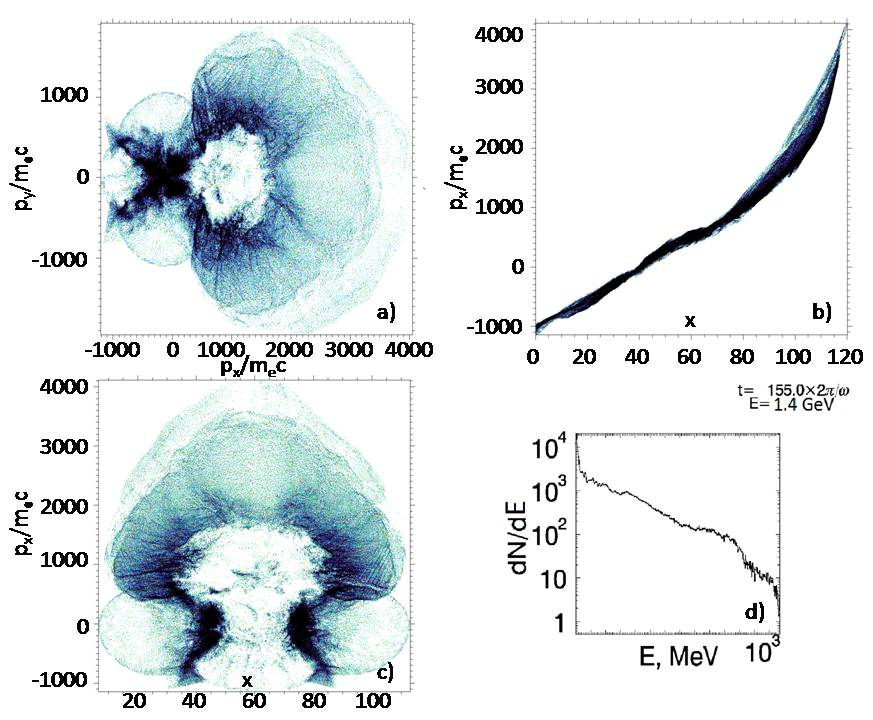}
\caption{The proton phase plots a) $(p_x,p_y)$, b) $(x,p_x)$, c) $(y,p_x)$, 
and d) the energy spectrum for $t=155$, $a_0  = 134$ in the case of normal laser incidence on the target.}
\label{ph134p}
\end{figure}

In the limit of the ultrahigh laser powers ($\approx 10 ~ {\rm PW}$) 
we can observe that these targets along with oblique incidence 
do not allow to reach the highest energies. Using the scaling parameter 
$a_0 = \varepsilon=2\pi n_e e^2 l/m_e \omega c$ (see Refs. \cite{VS, TZ}), 
we derive that the maximum energy 
of the ions is below $1~\rm GeV$, while for normal incidence and thicker 
central part of the target we can get energies larger at least by a factor 
of two. Here two mechanisms come into play. The first of them is the RPDA, 
when the laser pulse pressure wipes out the central--the densest part 
of the target and after that, 
it undergoes a Coulomb explosion according to Ref. \cite{DCE}, 
reaching the maximum energy of protons above $2~\rm GeV$ and $6~\rm GeV$ for oxygen ions for the target with $n_{\rm max}/n_{\rm cr}=80$. Using simple estimates for maximum ion energies gained in RPDA regime (\cite{DCE}, eq. (3)), we get lower values than observed in simulations, which could be caused by both Direct Coulomb Explosion and self-focusing of the laser pulse.
The parts of the target that were not in touch with 
the main pulse exploded after the squeezing by the magnetic field. 
Fig. \ref{fig14} shows how the magnetic field evolves in the case of 
$a_0=134$ and $n_{\rm max}/n_{\rm cr}=80$. 
Being focused into the preplasma, the maximum field amplitude increases 
at least by the factor of 1.5. 
Phase space distributions for ions can be seen on Fig. \ref{ph134p}. 
It is noticeable that magnetic squeezing effect 
still takes place for the ions at high laser pulse intensities in the regions of the target, 
which are not in the direct contact with the laser pulse 
(see Fig. \ref{ph134p}, hemispheres at $(p_x, p_y)$ and $(y,p_x)$ plots). 
Even though it did not give the most energetic ions, it still allows 
to reach $500~\rm MeV$ level.

Scaling of the maximum ion energy with the peak laser power 
in the log-log scale is presented in Fig. \ref{scaling}. 
The simulations of different targets (with various corona configurations) 
are presented as color dots, 
the line represents the expected maximum energy from the simple analytical 
estimation from the typical maximum energies 
obtained from Coulomb explosion, see Eqn. 1 from \cite{PREPLASMA}. 
Even though the maximum energy scales with the laser power ${\cal E}$ as 
$\sim {\cal P}^{1/2}$, we still obtain the ion energies that 
are high enough for the moderate laser systems. For instance, 
using these corona-equipped targets, 
we can reach the proton energies up to $200~ {\rm MeV}$ which is of high importance 
for the applications in hadron therapy \cite{HTUFN}.  
The oxygen ion energy is about $15~{\rm MeV/u}$ 
with the use of the finite contrast $200~ {\rm TW}$ laser systems. 
In the limit of Petawatt laser systems ($1-10~ {\rm PW}$), 
which would be accessible in the nearest future at ELI-Beamlines project \cite{ELI-BL}, 
we observe the shift to the foil acceleration by the 
radiation pressure rather than magnetic field amplified Coulomb explosion.

\begin{figure}
\includegraphics[width=8cm]{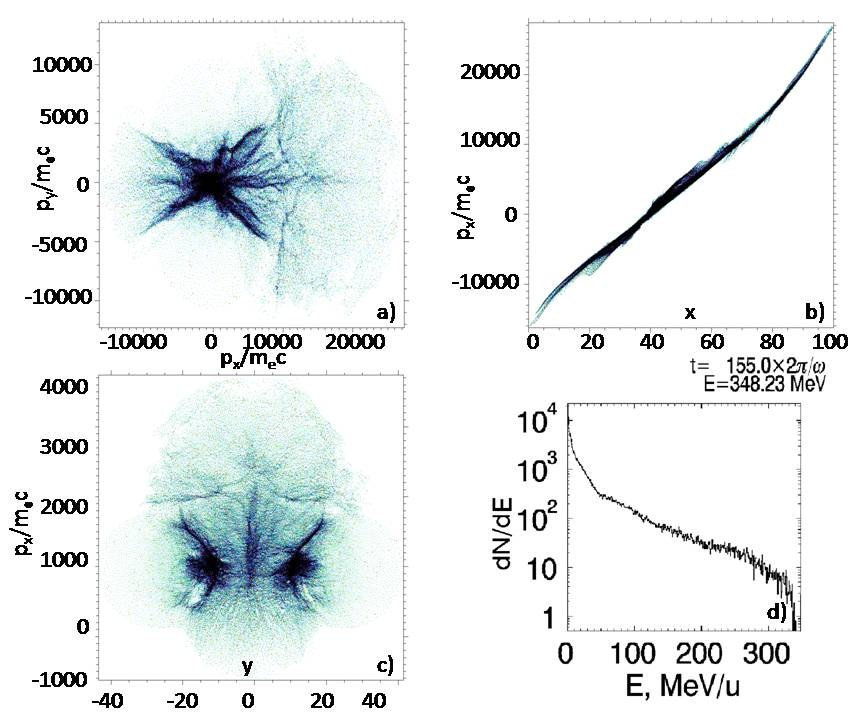}
\caption{The oxygen ion phase plots a) $(p_x,p_y)$, b) $(x,p_x)$, c) $(y,p_x)$, 
and d) the energy spectrum, $a_0  = 134$ in the case of normal laser incidence on the target.}
\label{ph134ox}
\end{figure}

\begin{figure}
\includegraphics[width=8cm]{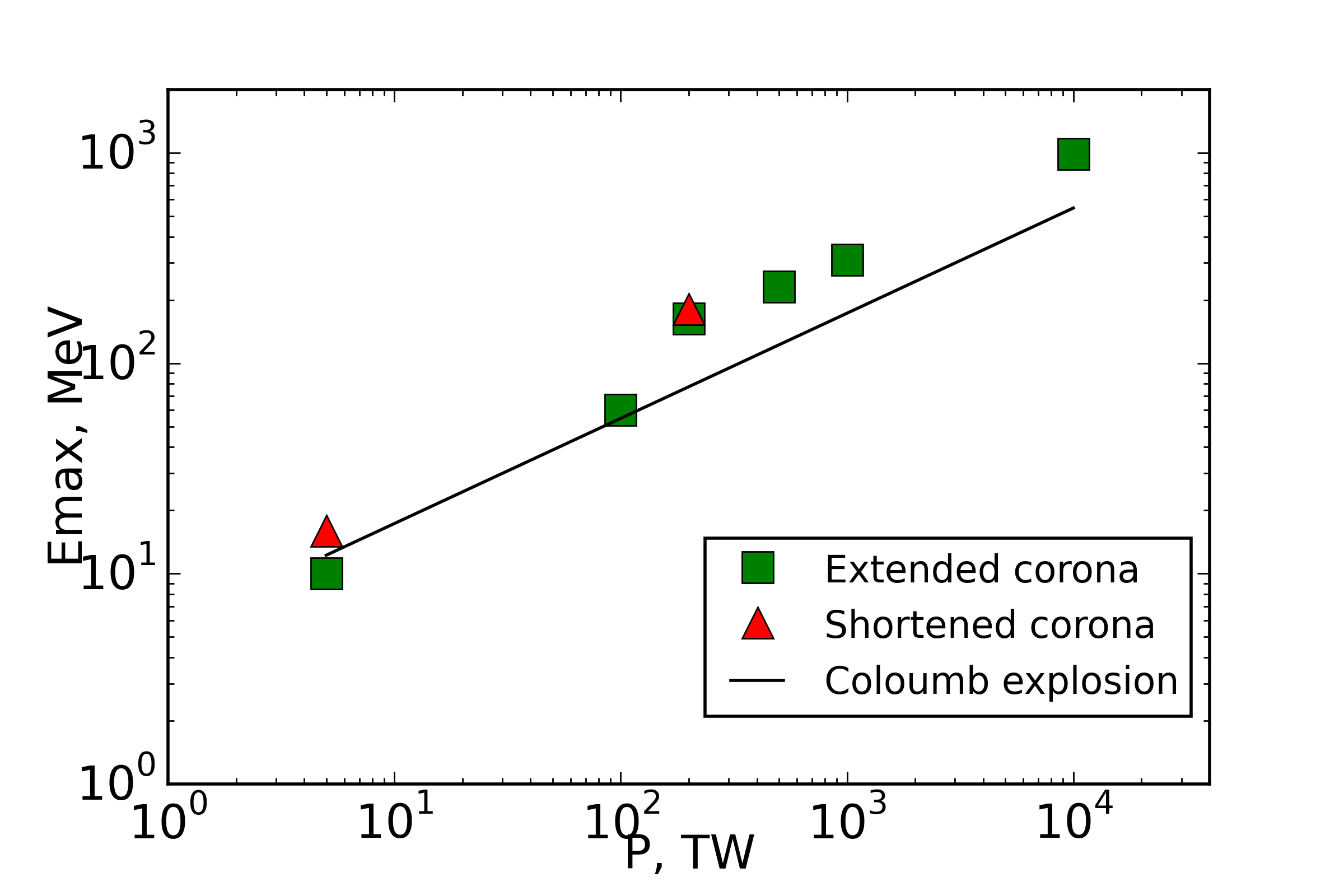}
\caption {Maximum proton energy at the peak of the laser pulse power: 
simulations with extended and shortened (exponential) corona vs 
theoretical curve for the Coulomb explosion (${\cal E}_p \approx 173\sqrt{{\cal P}[ \rm petawatt]}$ MeV, see \cite{PREPLASMA}).
}
\label{scaling}
\end{figure}

\section{Conclusion}

Here by using the 2D PIC simulations we show that the 
configurations corresponding to the mass-limited targets 
surrounded by an underdense plasma corona 
exhibit a number of effects enhancing the maximum energy 
 and increasing the total number of high energy ions. 
We observe the magnetic squeezing of the underdense parts of the target, 
when the pressure of self-generated 
quasi-static magnetic field pushes the electrons and ions into the 
target thus increasing the target plasma density.
This effect has an important implication at the stage of the 
Coulomb explosion enlarging the achievable ion energy.
While not giving the maximum energy in the whole fast ion population, 
it is still a major factor in the proton acceleration 
as it allows to accelerate a substantial (large compared with the 
maximum energy isotropic component) number of protons up 
to energies that are an order of magnitude larger than the majority of 
protons are accelerated to. 

Hole boring by the electric field arising due to the edge intensification 
and the plasma resonance at the sharp edge of the target is 
considered to be in strong correlation with the proton acceleration 
at the beginning of the interaction 
(before Coulomb explosions of the main target and preliminary compressed 
filaments come into play), 
providing the maximum proton energy at this interaction stage. 
Strong anisotropy of the proton 
acceleration in such a high-eccentricity ellipsoidal target 
is also an important feature that can be used 
for the generation of strongly collimated proton beams.

We also demonstrate than the presence of down-ramp density corona 
around the target can increase
 the maximum ion energies in comparison with flat density profiles. 

\section{Acknowledgements}

The authors thank M. Botton and A. Zigler for fruitful discussions. 
KVL expresses gratitude to M. Bussmann for the detailed comments on the 
early version of the work.
This work was supported by the ELI Project No. $\rm CZ.02.1.01/0.0/0.0/15\_008/0000162$. 
We also would like to acknowledge the support from Russian Foundation 
for Basic Research (grant No.~15-02-03063).

\end{document}